\pdfoutput=1
\documentclass{JINST}
\usepackage{subfigure}

\title{Study of Resistive Micromegas in a Mixed Neutron and Photon Radiation Field}

\author{Theodoros Alexopoulos$^a$, Georgios Iakovidis$^a$ $^b$\thanks{Corresponding
author.}~,  Georgios Tsipolitis$^a$\\
\llap{$^a$}National Technical University of Athens\\
  Zografou Campus, GR15773, Athens, Greece\\
\llap{$^b$}Brookhaven National Laboratory\\
  Upton, NY 11973, USA\\
  E-mail: \email{george.iakovidis@cern.ch}}

\abstract{The Muon ATLAS Micromegas Activity (MAMMA) focuses on the development and testing of large-area muon detectors based on the bulk-Micromegas technology. These detectors are candidates for the upgrade of the ATLAS Muon System in view of the luminosity upgrade of Large Hadron Collider at CERN (sLHC). They will combine trigger and precision measurement capability in a single device. A novel protection scheme using resistive strips above the readout electrode has been developed. The response and sparking properties of resistive Micromegas detectors were successfully tested in a mixed (neutron and gamma) high radiation field supplied by the Tandem accelerator, at the N.C.S.R. Demokritos in Athens. Monte-Carlo studies have been employed to study the effect of 5.5 MeV neutrons impinging on Micromegas detectors. The response of the Micromegas detectors on the photons originating from the inevitable neutron inelastic scattering on the surrounding materials of the experimental facility was also studied.}

\keywords{Micromegas; Gaseous Detectors}

\begin{document}

\section{Introduction}
The Micromegas (Micro-MEsh Gaseous Structure) detectors have been invented for the detection of ionizing particles in experimental physics, in particular in particle and nuclear physics. It was first proposed in 1996 \cite{Mic1}; its basic operation principle is illustrated in Figure 1. A planar drift electrode is placed few mm above a readout electrode. The gap is filled with ionization gas. In addition, a metal mesh is placed $\sim0.1$ mm above the readout electrode. The region between drift electrode and mesh is called the drift region, while the region between mesh and readout electrodes is called the amplification region. Both the mesh and the drift electrode are set at negative high voltage, so that a electric field of $\sim600$ V/cm is present in the drift region and a field of $\sim50$ kV/cm is present in the amplification region. The readout electrodes are set to ground potential through the pre-amplifier.
Charged particles transversing the drift region ionize the gas. The resulting ionization electrons drift towards the mesh with a typical drift velocity of  $\sim5$ cm/$\mu$s. The mesh itself appears transparent to the ionization electrons when the electric field in the amplification region is much larger than that in the drift region. Once reaching the amplification region, the ionization electrons cause a cascade of secondary electrons (avalanche) leading to a large amplification factor, which can be measured by the readout electrodes. A significant step in the development of Micromegas detectors was achieved in 2006 and its known as bulk-Micromegas technology. A detailed description can be found in \cite{Mic2}.
\section{Resistive Micromegas Chambers}
The resistive Micromegas developed by MAMMA group  \cite{Mic3}; has separate resistive strips rather than a continuous resistive layer to avoid charge spreading across several readout strips and to keep the area affected by a discharge as small as possible. The resistive strips are separated by an insulating layer from the readout strips and individually grounded through a large resistance. The Micromegas structure is built on top of the resistive strips. It employs a woven stainless steel mesh  which is kept at a distance of 128\,$\mu$m. from the resistive strips by means of small pillars. Above the amplification mesh, at a distance of 5\,mm, another stainless steel mesh  serves as drift electrode. Its lateral dimensions are the same as for the amplification mesh.The detectors, during the tests, have been operated with three Ar:CO$_{2}$  gas mixtures, with 80:20, 85:15 and 93:7.

\begin{figure}
\centering
\mbox{\subfigure{\includegraphics[width=2in]{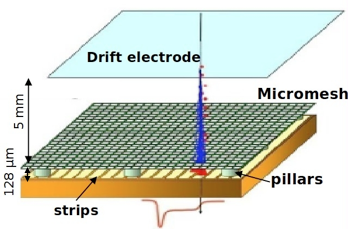}
\quad
\subfigure{\includegraphics[width=3.6in]{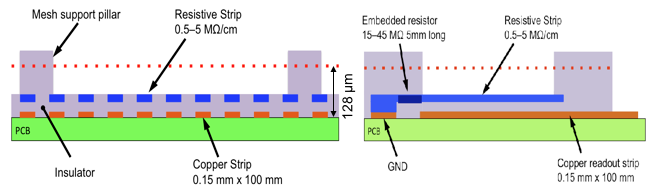} }}}
\caption{Resistive Micrommegas Layout.}
\end{figure}

\section{Neutron Test Facility}
 The mono-energetic neutron beam was produced via the $^2$H(d,$\textit{n}$)$^3$He reaction by bombarding a deuterium gas target with a deuteron beam at currents $\sim {1-2  \,\mu A}$. The deuteron beam was provided by the 5.5 MV HV TN-11 tandem accelerator of N.C.S.R. ``Demokritos'' in Athens, Greece. The gas target was fitted with a 5$\,\mu$s molybdenum entrance foil and a 1$\,$mm Pt beam stop. This was constantly cooled with a cold air jet during the irradiation to diminish the risk of damage to the Mo foil. The deuterium pressure was set to 1500$\,$mbar and was constantly monitored. Using this setup, the achieved flux varied between 3$\times$10$^5$ and 3$\times$10$^6\,\textit{n}/cm^2/$s at the Micromegas detector front, pending on the impinging beam current, measured directly on the deuterium gas cell.  The detector was placed at 0$^{\mathrm{o}}$ with respect to the neutron beam and at a distance of $\sim$50$\,cm$ from the center of the gas cell, thus limiting its angular acceptance to $\pm 5^{\mathrm{o}}$, in order to achieve the minimum possible uncertainty in the neutron beam energy. 

\subsection{Beam Properties}
Although the $^2$H(d,\textit{n})$^3$He reaction can produce mono-energetic neutrons up to 7.5$\,MeV$, it was shown with a BC501A liquid scintillator that there is an additional flux of parasitic low energy neutrons, originating mainly from (d,$\textit{n}$) reactions on low-$\textit{Z}$ impurities on the collimators and in the gas cell structural materials. This ``parasitic'' $\textit{n}$ flux increases with energy as shown in Figure 2. Below E$_{n,lab}$=7.5\,MeV its contribution remains lower than $\sim{20}$. The \textit{n} beam energy  used, was 5.5\,MeV since the number of parasitic \textit{n} is minimal also the differential cross-section of the $^{2}$H(d,\textit{n})$^3$He  reaction at 0$^{\mathrm{o}}$ remains adequate to have high \textit{n}-flux.
\begin{figure} [!htp]
	\begin{center}
		\includegraphics[scale=0.4]{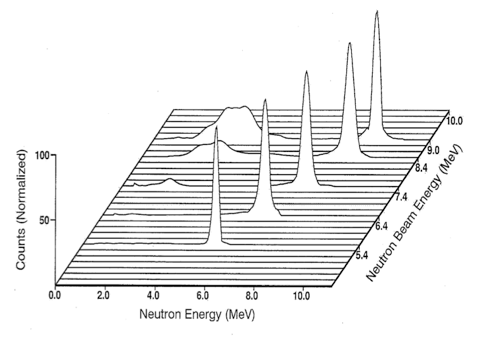}
		\caption{Correlation of ``parasitic'' \textit{n} flux  and \textit{n} beam energy.}
	\end{center}
\end{figure}

\subsection{Measurement of Neutron Flux}
The absolute neutron flux was measured experimentally with the method of activation of a thin $^{115}$In Foil  through the reaction $^{115}$In + $\textit{n} \rightarrow$   $^{115m}$In + $\gamma $ where the produced $^{115m}$In has half life time 4.486$\,h$ where they emit photons of energy 336.24$\,keV$. The produced $\gamma$  are detected through Ge $\gamma $-detector as shown in Figure 3.
\begin{figure} [!htp]
	\begin{center}
		\includegraphics[scale=0.4]{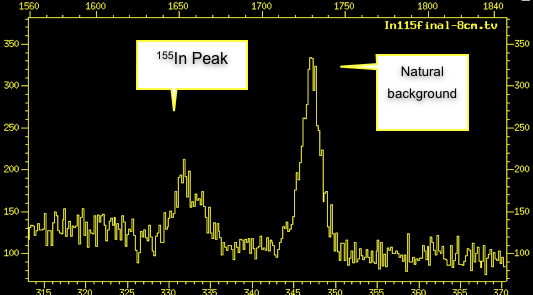}
		\caption{Neutron Flux measurement through the $^{115}$In + $\textit{n} \rightarrow$   $^{115m}$In + $\gamma $ reaction.}
	\end{center}
\end{figure}

\section{Monte Carlo}
The general purpose Monte Carlo code FLUKA \cite{Mic5} was used to study the effect of 5.5\,MeV \textit{n} impinging on a Micromegas detector. The response of the detector on the photons originating from the inevitable \textit{n} inelastic scattering on the surrounding materials of the experimental facility was simulated. To study the neutrons below 20\,MeV an additional routine was implemented, which explicitly takes into account the angular probability distribution functions for neutron elastic and inelastic scattering on $^{40}$Ar, $^{12}$C and $^{16}$O for E$_{\textit{n}}$ = 5.5\,MeV, as well as, detailed 2-body kinematics.  In the produced spectra the predominance of the \textit{p} induced signals is evident, the contribution of all the components of the gas mixture is illustrated (Figure 4). Secondary particles, such as electrons (from ionisation processes) and photons, depositing energy in  the active area of the detector, are  also   taken   into account.
\begin{figure} [!htp]
	\begin{center}
		\includegraphics[scale=0.38]{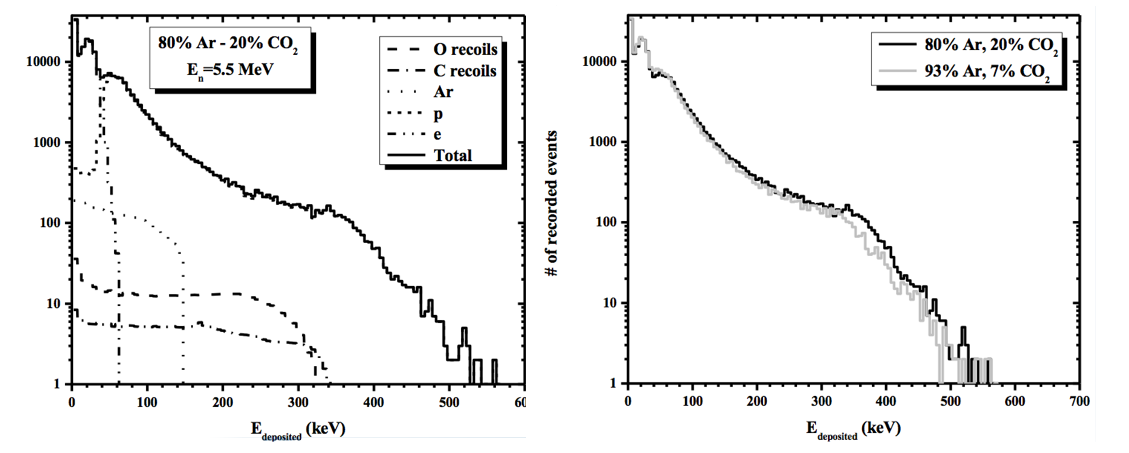}
		\caption{Left: Detailed simulation of the Micromegas detector's  response to 5.5\,MeV neutrons using FLUKA, along with the external algorithm for the heavy ion recoils. Right: Comparison of the Micromegas detector's  response to 5.5\,MeV neutrons, using two different gas mixtures, namely 80:20 and 93:7 Ar:CO$_{2}$.}
	\end{center}
\end{figure}

\section{Micromegas Performance in Neutron Beam}
We present here preliminary results on the performance of spark resistant Micromegas chambers in a beam of neutrons with a flux of $10^6\,\mathrm{Hz/cm^{2}}$. The detectors have been operated with three Ar:CO$_2$ gas mixtures, with 80:20, 85:15 and 93:7. Figure 5 shows a comparison of the the high voltage drop in case of sparks and the current that chamber draws for the bulk Micromegas on the left and resistive one on the right.  

\begin{figure} [!htp]
	\begin{center}
		\includegraphics[scale=0.45]{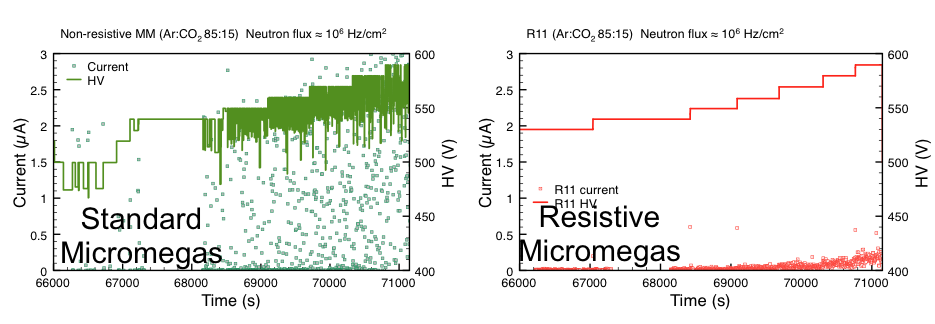}
		\caption{Micromegas HV drop in case of sparks. Left: Bulk Micromegas response. Right: Response of a resistive one.}
	\end{center}
\end{figure}

Only a few sparks per second were observed in a beam of 1.5$\times$10$^6$ \textit{n}$/$cm$^2/$s. The spark signal is reduced by a factor of 1000 compared to a standard Micromegas. The spark rate was found four times higher with the 80:20 compared to the 93:7 Ar:CO$_2$ gas mixture due to different transverse diffusion of drift electrons. The neutron interaction rate was found independent of the gas.

\acknowledgments
We would like to thank the MAMMA group at CERN for the support and development of Micromegas detectors as well as the ``Demokritos'' laboratory in Athens for providing the infrastructure for such an experiment.


\begin{thebibliography}{9}
\bibitem{Mic1} I. Giomataris, \textit{et al}., Micro-Pattern Gaseous Detectors, Nucl. Instrum. Meth. A 376 (1996) 29.
\bibitem{Mic2} I. Giomataris, \textit{et al}., Micromegas in a bulk; Nucl. Instrum. Meth., A 560 (2006) 405.
\bibitem{Mic3} Alexopoulos, T., \textit{et al}., A spark-resistant bulk-Micromegas chamber for high-rate applications, Nucl.Instrum.Meth., A 640 (2011) 110.
\bibitem{Mic4} Alexopoulos, T., \textit{et al}., Development of large size Micromegas detector for the upgrade of the ATLAS muon system, Nucl. Instrum. Meth., A 617 (2010) 161.
\bibitem{Mic5} Alexopoulos, T., \textit{et al}., Study of a Micromegas Chamber in a Mixed Neutron and Photon Radiation Field Using FLUKA, submitted to Nucl. Instrum. Meth., Nov. 2011.
\end{thebibliography}
\end{document}